# Article title: Population recordings of human motor units often display 'onion skin' discharge patterns - implications for voluntary motor control

Authors: Gregory EP Pearcey[1,2,3] and William Z Rymer[2,3]*

Affiliations: [1]School of Human Kinetics and Recreation, Memorial University of Newfoundland, St. John's, NL, Canada; [2]Shirley Ryan AbilityLab, and [3]Department of Physical Medicine and Rehabilitation, Feinberg School of Medicine, Northwestern University, Chicago, IL, USA

Number of words in abstract: 226
Number of words in manuscript: 4339

Address for correspondence:
W Zev Rymer
Shirley Ryan AbilityLab
355 E Erie Street, Chicago, IL 60611, USA.
Tel: +1 312 238 3919
E-mail: w-rymer@northwestern.edu

ABSTRACT

**Population recordings of human motor units display 'onion skin' discharge patterns - implications for voluntary motor control.**

Over the past two decades, there has been a radical transformation in our ability to extract useful biological signals from the surface electromyogram (EMG). Advances in EMG electrode design and signal processing techniques have resulted in an extraordinary capacity to identify motor unit spike trains from the surface of a muscle. These EMG grid, or high-density surface EMG (HDsEMG), recordings now provide accurate depictions of as many as 20-30 motor unit spike trains simultaneously during isometric contractions, even at high forces. Such multi-unit recordings often display an unexpected feature known as 'onion skin' behavior, in which multiple motor unit spike trains show essentially parallel and organized increases in discharge rate with increases in voluntary force, such that the earliest recruited units reach the highest discharge rates, while higher threshold units display more modest rate increases. This sequence results in an orderly pattern of discharge resembling the layers of an onion, in which discharge rate trajectories stay largely parallel and rarely cross. Our objective in this review is to explain why this pattern of discharge rates is unexpected, why it does not accurately reflect our current understanding of motoneuron electrophysiology, and why it may potentially lead to unpredicted disruption in muscle force generation. This review is aimed at the practicing clinician, or the clinician scientist. More advanced descriptions of potential electrophysiological mechanisms associated with 'onion skin' characteristics targeting the research scientist will be provided as reference material.

*Introduction*

Motor units (MUs) are the quantal elements required for regulating movement[1] and since select sets of muscle fibers are innervated by a single motoneuron (MN), spike times of the MU recorded from muscle fiber electrical activity (i.e the electromyogram; EMG) can provide information about the discharge of individual MNs in the central nervous system; the final common pathway for all motor behavior[2,3]. We also now know that the two ways of increasing force are:

1) activation of previously quiescent MUs or, put simply, recruitment; and
2) increases in discharge rate of already active units, known as rate modulation[4].

We now also know that during increasing excitation, MNs of smaller size are recruited before MNs of larger size. This concept, known as the ***size principle***[5–7], has been well-studied and has stood the test of time in both human and non-human animal models, with very few exceptions to date[8].

Once recruited, the discharge rate of a MN increases with increasing net excitatory synaptic input to the MN pool. Collectively, evidence gleaned using intracellular current injections during intracellular recordings indicates that MNs recruited with higher currents have the capacity to discharge at higher discharge rates than smaller, lower threshold MNs. Furthermore, when these discharge rate properties are assessed alongside data depicting the magnitude and time-course of the relevant MU mechanical twitches across recruitment thresholds, it seems appropriate that MU discharge rates should <u>increase</u> with increasing MN current threshold, since the rate of discharge required to reach a fused tetanic contraction is significantly higher for larger, faster, and stronger units than for their smaller, slower, and weaker counterparts. In other words, since further increases in discharge amount to very little extra force once slower units reach their fusion rate, it makes good sense that MUs made-up of slow-twitch muscle fibers require slow discharge rates to achieve a smooth tetanic contraction, when compared with MUs made-up of fast twitch muscle fibers. The relation between discharge rate and force development of the MU is summarized in figure 1 which shows that fusion of MU force requires higher discharge rates for high threshold MUs (red) than for low threshold units (blue).

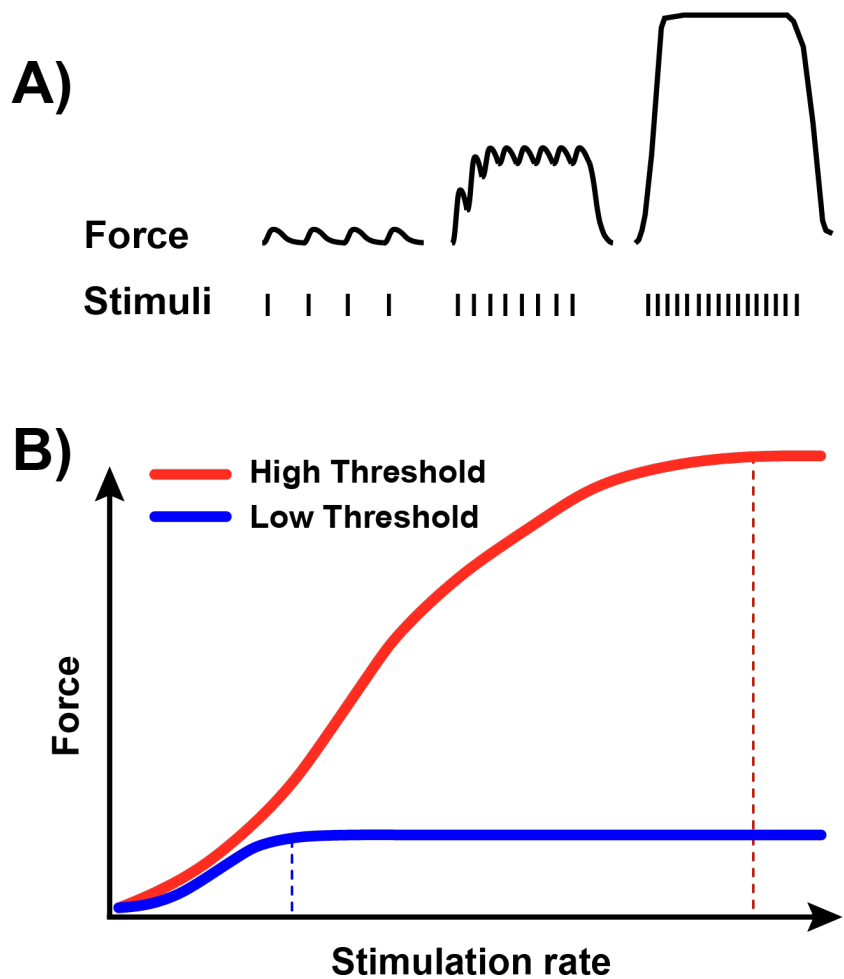

*Figure 1: Force output increases with a higher stimulus rate to the motor unit (A). This relationship is shifted to the right for high threshold (i.e., fast twitch) motor units, compared to lower threshold (i.e., slow twitch) motor units. This means that higher rates of stimuli are required to achieve maximal force from higher threshold motor units (vertical dotted lines).*

    Unexpectedly, recordings of many concurrently active MUs recorded from human muscles during submaximal isometric voluntary contractions, using surface EMG grids and advanced decomposition techniques, show completely different relationships between the expected recruitment and discharge rate properties. Specifically, later recruited units (i.e., higher threshold units) discharge at distinctly lower discharge rates than units recruited early in the contraction (i.e., lower threshold units). This 'onion skin' behavior, initially described by De Luca and colleagues[9–11], has since been reported frequently during submaximal isometric contractions in many muscles, and by various research groups across the world. A clear example

of the 'onion skin' pattern is depicted in figure 2, a diagrammatic representation of grid recordings summarizing typical results from several publications.

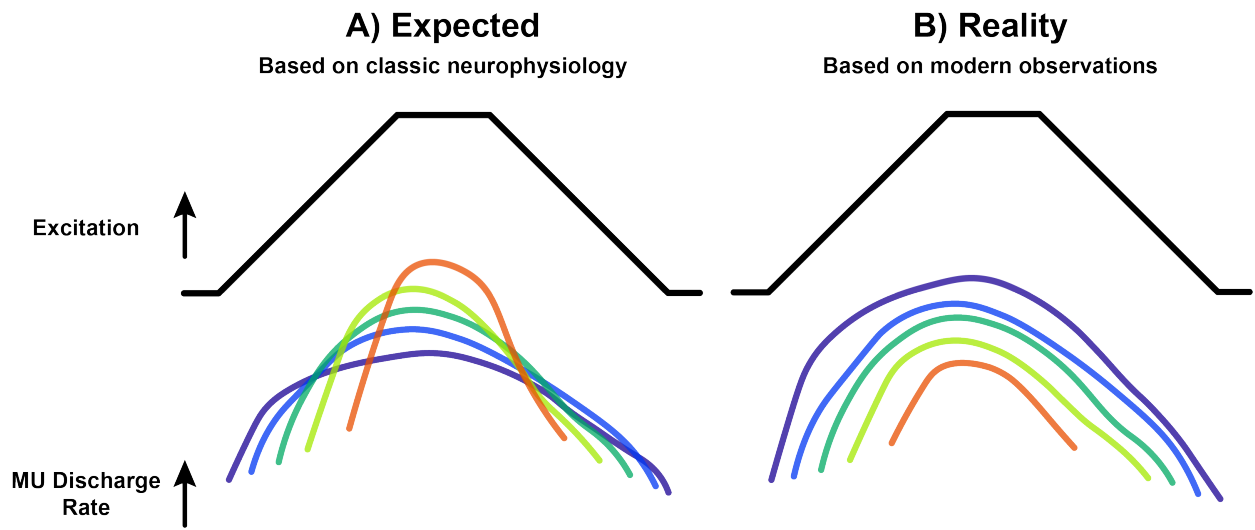

*Figure 2: Two simplified and theoretical depictions of motor unit behavior during a constant increase, hold, and decrease in excitation to submaximal levels. Two different possibilities of discharge rate profiles according to recruitment are shown. In A) the discharge rate of higher threshold units exceeds that of those recruited at lower thresholds, which is expected based on classic neurophysiological data examining the twitch and afterhyperpolarization durations in the reduced preparation. In B) the discharge rate of the earlier recruited units exceeds that of later recruited units, which only reach modest discharge rate increases – this is known as onion skin behavior.*

As such, we think it is now important to examine the potential underlying neural mechanisms that could cause such 'onion skin' behavior, and to also examine the functional implications of this behavior on motor control in health and disease.

### *Recruitment of motoneurons follows the 'size principle'*

Many MNs (often a few hundred) innervate a given muscle, and this group of neurons comprises the MN pool. Contemporary views even suggest that MN pools can span multiple muscles allowing for the coordinated control of synergistic actions (for examples see the work of Hug and colleagues[12,13]). Activity of the MN pool collectively modulates the force output from each muscle or groups of muscles, and MNs within the pool are activated progressively with increasing excitation, typically in order of the size and strength of the MU. The key biophysical properties of membrane resistance and MN surface area appear to dictate the net electrical input resistance of a MN[1], which ultimately governs the threshold for action potential initiation[14]. Essentially, when equal levels of synaptic input are provided to the entire motor pool, MNs with higher input resistance (i.e., smaller MNs) are recruited first, followed by MNs with decreasing input resistance (i.e., increasing size).

The idea of a 'size principle' or 'rank order of recruitment' was first advanced by Henneman in the mid 1950's using recordings from ventral root filaments and showing that motor axons of increasing diameter, identified using the size of the recorded action potential, were recruited progressively with stimulation to Ia afferents of increasing intensity[5]. In subsequent work, he and his colleagues substantiated their analysis by showing that MN size was directly related to axonal conduction velocity, which provides further support for an orderly rank order of recruitment[15], and this rank order was not altered by concurrent inhibitory inputs[16].

A short time later, Stein and colleagues showed that the size principle is preserved in humans by using the spike triggered averaging of individual MU recordings to estimate the twitch size of newly recruited units[17]. Since then, work from Cope and colleagues[8] has provided continuing and robust support for the generality of the size principle in mammalian preparations. In the decerebrate cat, for example, they showed that irrespective of the type of synaptic input (i.e., homonymous stretch, synergist stretch, or cutaneous afferent stimulation) recruitment was determined predominantly by cell size, with only occasional shifts in recruitment between units of similar properties. They concluded that these modest and occasional shifts in recruitment order are not likely to be functionally important[18,19]. Interestingly, Burke[20] has suggested that shifts in recruitment order probably require polysynaptic pathways and are highly state-dependent. Indeed, a realistic simulation from Heckman and Binder[21] showed that modifying the recruitment order requires a synaptic input that strongly favors high threshold MNs with a background level of inhibition to the motor pool. Although the order of recruitment may vary slightly, it has stood the test of time in that, typically, smaller MUs are recruited before larger ones. Discrepancies do however tend to emerge when the discharge rate of multiple units are tracked and compared when measured concurrently.

*Motoneuron discharge rates in relation to recruitment thresholds across the motor pool*

A single MN receives convergent synaptic input from thousands of other neurons that summate to create a net postsynaptic current, or a driving current and if that current is sufficiently large in amplitude and duration, repetitive discharge of the MN ensues[22]. To study such phenomena, intracellular recordings are the foundation upon which our knowledge of MN electrophysiology is built, because they allow recording of the membrane potential of the cell during spontaneous activation, or during synaptic input of different kinds[1].

Using intracellular recordings and current injections in anesthetized preparations, Eccles and colleagues[23] initially showed that MNs that fired slowly (i.e., so-called tonic MNs) had longer and deeper afterhyperpolarization potentials (AHP) than those that fired more rapidly (i.e., phasic MNs). Granit, Kernell and their colleagues[24–28] further showed that, in the anesthetized preparation, MN discharge frequency (f) is controlled chiefly by the intensity of maintained driving current (I), and this relation became known as the f-I relationship. The relationship between the MN discharge frequency and injected current intensity across a range of input currents is nearly linear, and the initial linear portion of this relationship is often labelled as the 'primary range.' A second, steeper but also near linear relationship between current and discharge rates, known as the 'secondary range,' can be achieved with higher intensity current injection. Although fundamental to our understanding of input-output properties of MNs, this 'base state' (i.e., deeply anesthetized) does not clearly reveal the actions of added neuromodulatory inputs that can modify these f-I relationships in awake and normal behaving mammals[29,30], an aspect that will be discussed later.

An accumulation of data from intracellular current injections suggests that the AHP is the primary determinant of the minimal discharge rate, since this rate is approximately equal to the reciprocal of the AHP duration, and AHP durations are reduced in MNs of increasing size. This led many to believe that MNs of increasing size should routinely discharge at higher rates because their AHPs are of shorter in duration[24]. Indeed, the discharge rate of newly recruited MNs was often shown to be greater than that of a lower threshold unit during synaptic activation in various types of non-human animal preparations (see fig 2 from Powers and colleagues[31]).

For example, during stimulation of the mesencephalic locomotor region (i.e., stimulation of the brainstem) the pattern of discharge rate modulation of MN pairs is related to their recruitment order, when assessed during smoothly graded contractions[32,33]. In general, lower threshold MUs had lower initial and peak discharge rates than higher threshold units. Some pairs of MUs that were recruited at similar thresholds increased their discharge rates similarly (i.e., nearly parallel increases in discharge rates) whereas MNs that were recruited with very large differences in recruitment thresholds did not modulate their discharge rates in a correlated fashion. Later work from Prather et al.[34] showed that, regardless of the type of synaptic input from afferent sources (i.e., cutaneous or muscle afferents), the discharge rate modulation of a higher threshold MN was typically higher than the modulation in discharge rate of a lower threshold MN. This type of behavior typically resulted in higher threshold MNs achieving higher discharge rates than the lower threshold MNs, despite some degree of co-modulation.

Based then on multiple studies from animal preparations, it seems that MN recruitment and discharge properties are likely to be well-matched to the mechanical behavior of the associated muscle units, in that the discharge properties of slow-twitch units are governed by the prolonged after hyperpolarization (AHP) of the associated MN and this matches the mechanical behavior of the associated muscle fibers. We next describe these relations between MNs and their associated muscle fibers in more detail.

***Twitch properties in relation to recruitment thresholds and discharge rates across the motor pool***

Discharge rates of different MN types throughout the motor pool were believed to match the twitch properties of the innervated muscle fibers, in that MNs innervating slow twitch fibers discharge more slowly than MNs innervating fast twitch fibers. The idea was first proposed at the level of a whole muscle, where certain muscles were considered to show fast contractions (i.e., pale muscle), whereas others were considered 'slow' (i.e., red muscle; see Ranvier[35] and Kronecker & Stirling[36]). Denny-Brown[37] characterized differences in these muscle types in the synergistic triceps surae based on their histological appearance. The soleus is comprised primarily of red muscle fibers activated in a tonic fashion, whereas the gastrocnemius had more pale muscle fibers with more phasic activation. This concept, however, was slightly flawed in that most muscles contain a mixture of different colored muscle fibers indicating a mixture of MU types within each muscle.

Regardless of the limitations, examination of antidromic action potentials monitored via intracellular recording of MNs innervating red (slow) and pale (fast) muscle fibers revealed that

MNs innervating slower muscle had longer AHP durations[23]. In fact, across the 26 muscles studied, the duration of AHP had an inverse relationship with the conduction velocity of the motor axons, which suggested that axons of smaller diameter were usually linked to slow MNs. These slow or 'tonic' MNs innervated muscles containing slow twitch fibers (i.e., red muscle) that required relatively low discharge rates to reach near maximum tetanus (i.e., fusion of twitches to produce near-maximal tension such that there is no reduction in tension between the twitches), when compared to fast or 'phasic' MNs. As such, Eccles and colleagues[23] suggested that "a high frequency discharge to slow Mus would merely serve to fatigue the muscle for no effective return in a higher contraction tension, while the low frequency discharge from a tonic motoneuron to a fast MU would be inefficient in fusing the individual twitch responses to give an effective tetanic contraction."

Furthering the idea of a useful and functional relation between the electrophysiological properties of MNs and the muscle fibers they innervate (i.e., as suggested first by Eccles and colleagues), Burke and his colleagues[38–42] classified MUs into distinct types. MUs were classified as either: type S (slow twitch, all resistant to fatigue); type FR (fast twitch, fatigue resistant); type FF (fast twitch, fatigable); or sometimes, type FI (fast twitch, intermediate fatigue; often unclassified) (see [43] for a historical perspective of the classification). The first recruited MNs (i.e. type S) innervated slow twitch muscle fibers, generated small amplitude twitches with slow decay of the twitch, were difficult to fatigue, and were able to sustain stable contractions for many minutes. Conversely, large MNs routinely innervated fast twitch muscle fibers that were able to generate large amplitude twitches, with rapid force decay, most of which were also highly fatigable (type FF). Some of these larger units were resistant to fatigue (type FR)[41].

The time course of twitch tensions provides particularly fruitful information about the optimal discharge rates of MUs of various sizes (see figure 3). Large twitches of MUs with higher thresholds are brief (i.e., faster times to peak and shorter relaxation times), whereas twitches of smaller MUs are typically more prolonged in both rising and falling phases[44]. As one might imagine, brief twitches require higher frequencies to achieve smooth contractions (i.e. fusion or tetanus), whereas prolonged twitches would require much lower frequencies to enable fused contractions. In line with these recorded differences in twitch amplitudes and durations, type S units have longer AHP durations than type F [45–49], suggesting that lower threshold MUs should discharge at lower rates than higher threshold units.

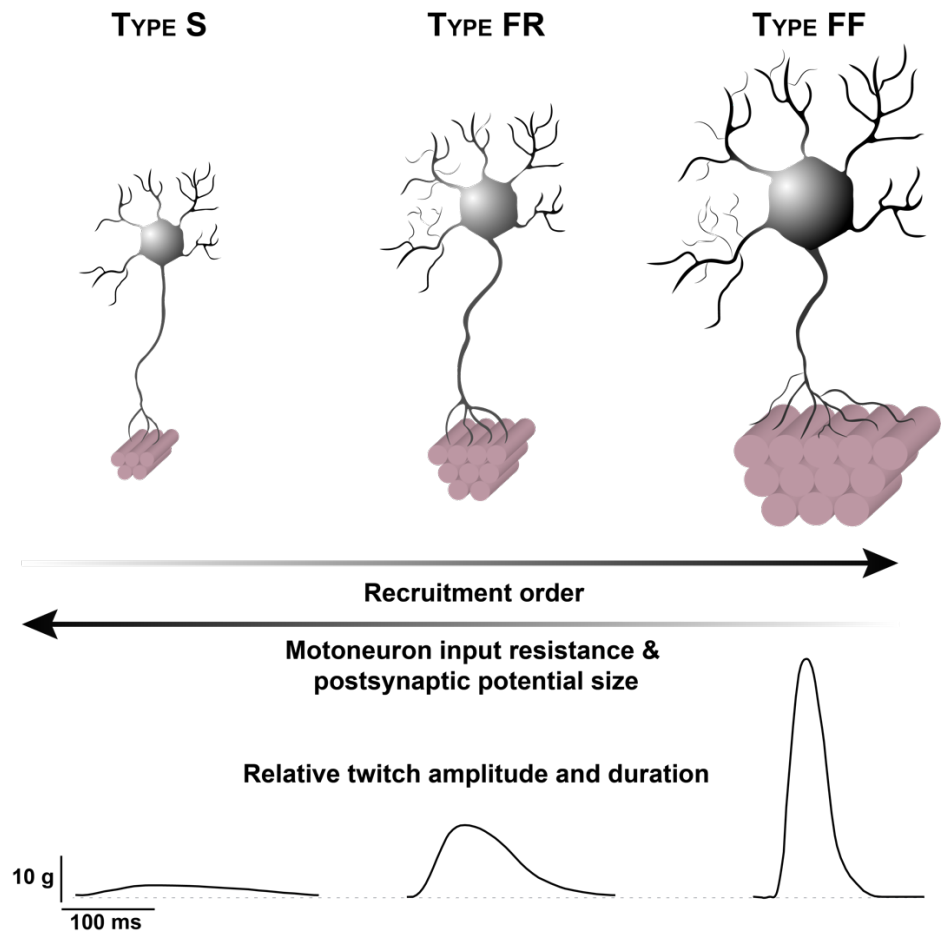

*Figure 3: A simplified depiction of motor unit types that comprise a continuum. Motor units are grouped according to their size, which primarily determines their recruitment order. As size of the parent motoneuron increases, so does the number and size of innervated muscle fibers. Motor units recruited first (i.e., type S) have the lowest twitch amplitudes with slow relaxation times, whereas those recruited later have twitches with substantially larger amplitudes but brief relaxation times.*

      Under steady-state conditions, lower threshold MUs do indeed reach their maximal force output (i.e. fusion rate) at lower discharge rates than higher threshold units[50,51], as shown in figure 1. This force-frequency relation describes the conversion of MN action potentials into isometric muscular force and provides the necessary link between the MN and its innervated muscle fibers[1]. Therefore, Kernell[52], in line with Eccles[23], suggested that since high discharge rates produce very little extra force output, they are energetically wasteful in low threshold MUs,

and therefore, one would expect the range of discharge rates to increase with MU size. This has been the prevailing thought amongst the MN field - that is, small MUs are recruited at low thresholds and should maintain relatively low discharge rates due to their long twitches and AHP durations compared to larger MUs recruited at higher thresholds with shorter twitches and AHP durations.

***Lessons gleaned from high-density electromyogram grids***

In the last two decades, there has been increasing use of high-density surface EMG (HDsEMG) recordings[53]. These HDsEMG grids have varied in size from a small number of electrodes (less than 10) to very large arrays with 128 or even 256 electrodes. Electrode dimensions have also varied but for the most part electrodes have retained relatively large dimensions (greater than 1 cm$^2$). These recent advances in electrode design, in addition to vast improvements in computational power allow for the recording of many human MUs simultaneously using HDsEMG. Particular credit is owed to development and improvements in blind source separation algorithms, developed by Holobar, Farina, Merletti, Negro and their colleagues[54–63]. As a result, we are now able to record MU spike times from the surface of muscles ranging in function and structure. HDsEMG grids have been used to record from small hand muscles, such as the first dorsal interosseous (FDI), or from large limb muscles such as vastus lateralis or biceps brachii, to reveal the spike times of as many as ~50 concurrently active MUs within a single trial. Unexpectedly, what we see in these recordings diverges substantially from findings derived from the aforementioned classic MN studies.

Although MUs appear to be recruited in accordance with their size (as estimated by MU action potential amplitude) the discharge rate behavior recorded using high density grids diverges radically from expectations based on classic MN studies. We would expect, from classical MN physiology summarized earlier, that earlier recruited MUs would discharge more slowly, presumably optimizing the metabolic and mechanical properties of the muscle fibers they innervate. Instead, 'onion skin' behavior becomes manifest in most human data (refer to figures 2 and 4). The first units to be recruited (blue units) show continuing increases in discharge rate with increasing force, yet later recruited units discharge at slower rates (green/yellow units), and the last recruited units discharge at the slowest rates (red units).

This orderly behavior of different MUs is quite unexpected and gives rise to a layering of discharge rate trajectories that rarely cross in HDsEMG recordings of neurologically intact humans during submaximal isometric contractions. Although there are occasional crossings in the discharge rate trajectories, most of these occur in units that are recruited at similar thresholds (i.e., neighboring blue units cross in figure 1). On the other hand, it is very rare for the discharge rates of MUs with substantial recruitment spacing to cross (i.e., red unit discharge rates do not cross those of blue units in figure 4).

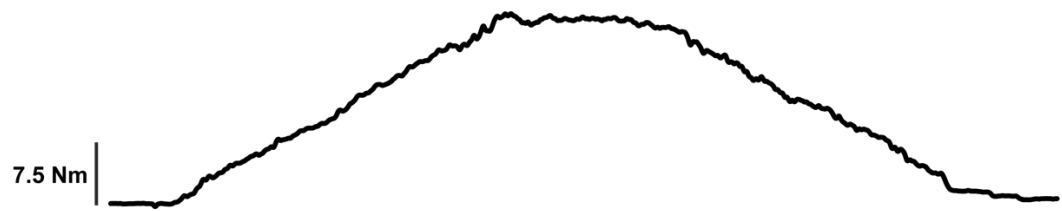

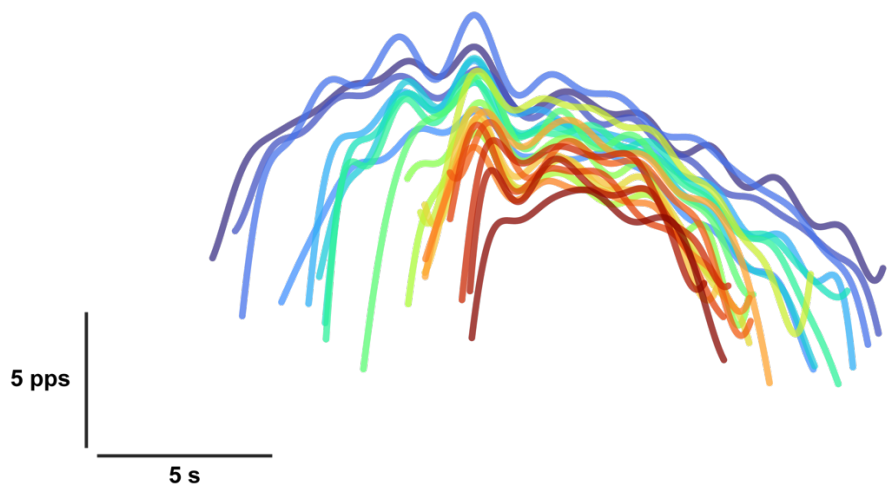

*Figure 4: The torque (A) and smoothed discharge behavior (B) of 20 motor units recorded simultaneously using high-density surface electromyography and blind source separation during an isometric dorsiflexion task to 30% of maximum. Note the discharge of the earliest recruited motor units (blue) reach much higher rates than the latest recruited units (red).*

### *When do 'onion skin' profiles emerge?*

Virtually all the published records of 'onion skin' profiles are based on decomposition of MU spike times from EMG grid recordings taken from the skin surface over a muscle during increases in isometric voluntary force (for examples see [64,65]) or during submaximal voluntary efforts during slow shortening or lengthening of the muscle[66], although records of this behavior began with intramuscular EMG recordings in the 1970's[67,68]. To date, there are almost no other recordings illustrating the behavior of several MUs in which activation is mediated by other kinds of synaptic input. In particular, we have almost no recordings in which many units are tracked and recruited by muscle afferent excitation[34].

There are several hypothetical mechanisms that could be considered when trying to understand or explain these 'onion skin' discharge rate trajectories. For example, the electrical impedance of spinal MNs (related to the electrical resistance) could be modified during voluntary contractions so that higher threshold units show disproportionate reductions in membrane electrical impedance during progressive MN recruitment. This could potentially happen because of concurrent inhibitory inputs that grow more rapidly than excitatory input during voluntary contractions. We do not yet know of any appropriate circuits that may reveal these properties, in which inhibition of MNs increases disproportionately, but this remains a plausible possibility[29,69,70].

Alternatively, voluntary commands could harness other sources of disruptive inhibitory synaptic input, acting potentially via regional spinal interneurons. This disruption could impact higher threshold MNs disproportionately and could include changes in presynaptic control that could regulate the magnitude and distribution of excitation to the motor pool.

***'Onion skin' behavior is not restricted to human preparations***

Although commonly assumed to be a phenomenon manifest exclusive to human preparations, 'onion skin' behavior has also been documented in non-human animal preparations too. Henneman and colleagues[71] showed that MNs recruited earlier during high frequency plantaris nerve stimulation have higher discharge rates than those recruited later (see fig. 6 from Henneman et al.[71]; panel A low ~11pps, high ~7pps; panel B low ~15pps, high ~12pps). Data obtained from Christopher Thompson and colleagues (unpublished Heckman Lab findings) recorded with HDsEMG over the medial gastrocnemius muscle of the decerebrate cat also display 'onion skin' behavior during progressive muscle stretch (see figure 5). This behavior is strikingly similar to that seen in human MU recordings during submaximal voluntary contractions where discharge rates are higher in low threshold MUs compared to the later recruited and higher threshold units (i.e., compare figures 4 and 5).

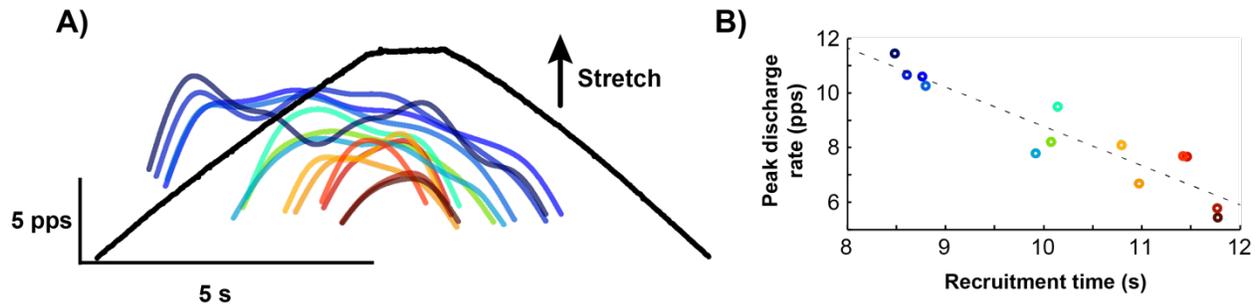

*Figure 5: Linear stretch of the isolated medial gastrocnemius in a decerebrate cat while EMG is recorded with 64 channel HDsEMG recording electrodes. Signals were then decomposed into corresponding motor unit action potentials using blind source separation and discharge rates smoothed with a 1s Hanning window. (A) An example stretch (5 mm; 1 mm/s; black trace) of the MG reveals the smoothed discharge rate of 13 motor units (colored traces). (B) A negative correlation is observed when the peak discharge rate is plotted against time of recruitment during the stretch for each of the 13 motor units, indicating the later recruited units discharged at lower rates when compared to earlier recruited units (i.e., 'onion skin').*

***Role of persistent inward currents (PICs) and direct actions of monoamines***

The contribution of anesthetized preparations to our understanding of input-output properties in MNs is fundamental, but without the effects of key metabotropic inputs originating from the brainstem (i.e. serotonin [5HT] or norepinephrine [NE]) we may not fully understand the way that MNs function during normal behavior. Passive neuronal properties alone cannot account for the MN recruitment and discharge rate modulation patterns seen routinely in awake behaving mammals[72]. Monoaminergic inputs must also be considered because of their profound direct effects on the excitability of spinal MNs[73], as well as their well-documented facilitation of dendritic persistent inward currents (PICs)[74–78], which are voltage-gated ion channels well-equipped to amplify and prolong excitatory synaptic inputs to the MN by providing a sustained depolarising current to the MNs[30,79].

***Persistent inward currents (PICs) and 'onion skin' discharge***

To the best of our knowledge, PICs themselves do not induce major changes in membrane electrical impedance, although this thesis has not been widely tested. One factor limiting our ability to understand this concept is that most studies of PICs examine those synaptic mechanisms which are mediated largely by excitable sodium channels, located on the MN soma. We are less able to extensively study dendritic PICs that are mediated primarily by calcium channels. It remains possible that since 1) sodium PICs are necessary for repetitive discharge [80]; and 2) self-sustained discharge, a hallmark of PICs, is longer-lasting in low-threshold MUs[81], PICs could be more effective in smaller compared to large MNs. Although necessarily speculative, this differential contribution of PICs to MNs of various thresholds could contribute to 'onion skin' behavior by reducing AHP disproportionately in small units with similar driving currents.

*Monoamines and 'onion skin' rate profiles*

We have known for many years that several monoamines including serotonin and epinephrine can modify ion channel conductances in a progressive and orderly way, so it is possible that these same may contribute to the 'onion skin' pattern. For example, it could be the case that increasing voluntary force increases the amount of a monoamine released in the spinal cord[82,83] and that this increase progressively exerts a differential effect on smaller versus larger MNs. This may be an unrealistic scheme of control because it requires fine regulation of the transmitter effects in the cord, a complex task for slowly discharging monoaminergic neurons in the brainstem.

**Functional consequences of 'onion skin' discharge rate profiles**

As we have described in earlier paragraphs, the parallel rate trajectories of successively recruited MUs were unexpected, given the well described relations between injected current and discharge rate of different sized MNs reported in the literature. Specifically, we would have expected discharge rates of higher threshold units to be systematically greater, to match the higher fusion frequency of their innervated faster-twitch muscle fibers. This should result in a radically different discharge rate picture, with higher threshold units showing steeper discharge rate trajectories as force increases, resulting in multiple discharge rate trajectory crossings. This was not the case, however, as has now been documented in many reports.

One unexpected and potentially adverse consequence of such a discharge scheme is that higher threshold units are now being activated with discharge rates that are much further from the optimal fusion frequency of their innervated muscle fibers. In other words, the force traces after recruitment of higher threshold units could potentially appear to be "bumpy" giving rise to greater force variability at higher muscle forces.

There are several lines of evidence, drawn from human behavioral studies, suggesting that force variability is indeed greater at higher force levels[84], as would be expected from these discharge schemes. There are direct measures of isometric force variation, such as those reported by Jones and colleagues[85] from Daniel Wolpert's group during contractions of arm and hand muscles. Others have used direct measures of force variability, such as the mechanomyogram (MMG) or the use of optical sensors, to track surface motion of a muscle and show similar features[86]. So, it does appear that low discharge rates of high threshold units may have demonstrable functional consequences.

It remains unclear what functional benefits could derive from such a scheme. Several years ago, De Luca and his colleagues[9,11], who first coined the phrase 'onion skin', suggested that it might constitute a form of functional reserve allowing fast twitch MUs to remain unfatigued and to be available for force generation during emergencies[11]. The notion of having MUs in reserve might seem like a bit of a stretch but limiting the use of the most fatigable and energetically demanding units (i.e., those of higher thresholds) until they are absolutely required may serve as a trade-off with the fluctuations in force to reduce fatigue during normal submaximal behaviors. Although these outlined benefits of the 'onion skin' discharge scheme remains plausible, it would seem to be a highly inefficient, in terms of force generation from an individual unit, in that MUs are being activated far from their optimal force generation capacity over much of their normal range of behaviors.

Given the recent and radical improvements in ultrasound technology, we may soon be able to track fascicle motion and force generation of muscle fascicles at different discharge rates and confirm some of these suppositions. At the moment, hypotheses about the functional benefits arising from such 'onion skin' discharge schemes remain necessarily speculative.

**Summary**

This review examines the impact of EMG recordings made using the relatively new technology of high-density EMG grids. These grids are able to record high resolution EMG signals from the skin overlying a muscle surface, and when combined with advanced signal processing techniques, we are now able to extract multiple MU spike trains simultaneously. It was widely expected that the resultant recordings would show that low threshold MUs would fire consistently low of discharge rates, and that high threshold units would show more rapid increases in discharge rate with increasing voluntary force commands. Unexpectedly, this did not happen and there are now multiple reports in which increases in force generation are accompanied by parallel increases in discharge rate of many MUs resulting in and 'onion skin' appearance.

We do not know the mechanisms of such a scheme of recruitment and rates increases yet, and we also do not know if there are any advantages of such a scheme. In fact, it seems likely that this 'onion skin' scheme results in greater force variability during high level contractions. There may also be even greater variability of force generation in neurological impairments if the 'onion skin' rate pattern is also prevalent. So far, the scheme also seems to be followed in muscles of individuals living with chronic stroke and spinal cord injury in whom discharge rates are often generally depressed. This may mean, in turn that variance of force generation is also greater in people living with neurological impairments.

**Acknowledgements**

The authors express their gratitude to CJ Heckman for helpful comments on a draft of the manuscript, and to Christopher K Thompson for providing unpublished motor unit data from a decerebrate cat experiment.

# References

1. Heckman CJ, Enoka RM. Motor unit. *Compr Physiol*: 2012;2:2629–2682. https://doi.org/10.1002/cphy.c100087.

2. Sherrington CS. Correlation of reflexes and the principle of the common path. *Nature*: 1904;70:460–466.

3. Charles S. Sherrington. *The Integrative Action of the Nervous System*. Yale University Press: New Haven, CT, 1906.

4. Adrian ED, Bronk DW. The discharge of impulses in motor nerve fibres: Part II. The frequency of discharge in reflex and voluntary contractions. *J Physiol*: 1929;67:i3-151.

5. Henneman E. Relation between Size of Neurons and Their Susceptibility to Discharge. *Science*: 1957;126:1345–1347.

6. Henneman E, Somjen G, Carpenter DO. FUNCTIONAL SIGNIFICANCE OF CELL SIZE IN SPINAL MOTONEURONS. *J Neurophysiol*: 1965;28:560–580. https://doi.org/10.1152/jn.1965.28.3.560.

7. Henneman E, Somjen G, Carpenter DO. Excitability and inhibitability of motoneurons of different sizes. *J Neurophysiol*: 1965;28:599–620. https://doi.org/10.1152/jn.1965.28.3.599.

8. Cope TC, Clark BD. Motor-unit recruitment in the decerebrate cat: several unit properties are equally good predictors of order. *J Neurophysiol*: 1991;66:1127–1138. https://doi.org/10.1152/jn.1991.66.4.1127.

9. De Luca CJ, Erim Z. Common drive of motor units in regulation of muscle force. *Trends Neurosci*: 1994;17:299–305. https://doi.org/10.1016/0166-2236(94)90064-7.

10. De Luca CJ, Hostage EC. Relationship between firing rate and recruitment threshold of motoneurons in voluntary isometric contractions. *J Neurophysiol*: 2010;104:1034–1046. https://doi.org/10.1152/jn.01018.2009.

11. De Luca CJ, Contessa P. Biomechanical benefits of the Onion-Skin motor unit control scheme. *J Biomech*: 2015;48:195–203. https://doi.org/10.1016/j.jbiomech.2014.12.003.

12. Hug F, Avrillon S, Sarcher A, Del Vecchio A, Farina D. Correlation networks of spinal motor neurons that innervate lower limb muscles during a multi-joint isometric task. *J Physiol*: June 2022. https://doi.org/10.1113/JP283040.

13. Hug F, Avrillon S, Ibáñez J, Farina D. Common Synaptic Input, Synergies, and Size Principle: Control of Spinal Motor Neurons for Movement Generation. July 2022. https://doi.org/10.48550/arXiv.2208.02818.

14. Henneman E, Mendell LM. Functional organization of motoneuron pool and its inputs. In: Handbook of Physiology, Section 1, The Nervous System. Vol 2. American Physiological Society: Bethesda, MD; 1981. p. 423–507

15. Clamann HP, Henneman E. Electrical measurement of axon diameter and its use in relating motoneuron size to critical firing level. *J Neurophysiol*: 1976;39:844–851. https://doi.org/10.1152/jn.1976.39.4.844.

16. Clamann HP, Gillies JD, Henneman E. Effects of inhibitory inputs on critical firing level and rank order of motoneurons. *J Neurophysiol*: 1974;37:1350–1360. https://doi.org/10.1152/jn.1974.37.6.1350.

17. Milner-Brown HS, Stein RB, Yemm R. The orderly recruitment of human motor units during voluntary isometric contractions. *J Physiol*: 1973;230:359–370. https://doi.org/10.1113/jphysiol.1973.sp010192.

18. Cope T, Pinter M. The Size Principle: Still Working After All These Years. *Physiology*: 1995;10:280–286. https://doi.org/10.1152/physiologyonline.1995.10.6.280.

19. Cope TC, Sokoloff AJ. Orderly recruitment tested across muscle boundaries. *Prog Brain Res*: 1999;123:177–190. https://doi.org/10.1016/s0079-6123(08)62855-1.

20. Burke RE. Some unresolved issues in motor unit research. *Adv Exp Med Biol*: 2002;508:171–178. https://doi.org/10.1007/978-1-4615-0713-0_20.

21. Heckman CJ, Binder MD. Computer simulations of the effects of different synaptic input systems on motor unit recruitment. *J Neurophysiol*: 1993;70:1827–1840. https://doi.org/10.1152/jn.1993.70.5.1827.

22. Powers RK, Binder MD. Input-output functions of mammalian motoneurons. *Rev Physiol Biochem Pharmacol*: 2001;143:137–263. https://doi.org/10.1007/BFb0115594.

23. Eccles JC, Eccles RM, Lundberg A. The action potentials of the alpha motoneurones supplying fast and slow muscles. *J Physiol*: 1958;142:275–291. https://doi.org/10.1113/jphysiol.1958.sp006015.

24. Kernell D. SYNAPTIC INFLUENCE ON THE REPETITIVE ACTIVITY ELICITED IN CAT LUMBOSACRAL MOTONEURONES BY LONG-LASTING INJECTED CURRENTS. *Acta Physiol Scand*: 1965;63:409–410. https://doi.org/10.1111/j.1748-1716.1965.tb04081.x.

25. Granit R, Kernell D, Shortess GK. QUANTITATIVE ASPECTS OF REPETITIVE FIRING OF MAMMALIAN MOTONEURONES, CAUSED BY INJECTED CURRENTS. *J Physiol*: 1963;168:911–931. https://doi.org/10.1113/jphysiol.1963.sp007230.

26. Granit R, Kernell D, Shortess GK. THE BEHAVIOUR OF MAMMALIAN MOTONEURONES DURING LONG-LASTING ORTHODROMIC, ANTIDROMIC

AND TRANS-MEMBRANE STIMULATION. *J Physiol*: 1963;169:743–754. https://doi.org/10.1113/jphysiol.1963.sp007293.

27. Granit R, Kernell D, Lamarre Y. Algebraical summation in synaptic activation of motoneurones firing within the "primary range" to injected currents. *J Physiol*: 1966;187:379–399. https://doi.org/10.1113/jphysiol.1966.sp008097.

28. Granit R, Kernell D, Lamarre Y. Synaptic stimulation superimposed on motoneurones firing in the "secondary range" to injected current. *J Physiol*: 1966;187:401–415. https://doi.org/10.1113/jphysiol.1966.sp008098.

29. Johnson MD, Thompson CK, Tysseling VM, Powers RK, Heckman CJ. The potential for understanding the synaptic organization of human motor commands via the firing patterns of motoneurons. *Journal of Neurophysiology*: 2017;118:520–531. https://doi.org/10.1152/jn.00018.2017.

30. Heckman CJ, Johnson M, Mottram C, Schuster J. Persistent inward currents in spinal motoneurons and their influence on human motoneuron firing patterns. *Neuroscientist*: 2008;14:264–275. https://doi.org/10.1177/1073858408314986.

31. Powers RK, Nardelli P, Cope TC. Estimation of the Contribution of Intrinsic Currents to Motoneuron Firing Based on Paired Motoneuron Discharge Records in the Decerebrate Cat. *Journal of Neurophysiology*: 2008;100:292–303. https://doi.org/10.1152/jn.90296.2008.

32. Tansey KE, Botterman BR. Activation of type-identified motor units during centrally evoked contractions in the cat medial gastrocnemius muscle. I. Motor-unit recruitment. *J Neurophysiol*: 1996;75:26–37. https://doi.org/10.1152/jn.1996.75.1.26.

33. Tansey KE, Botterman BR. Activation of type-identified motor units during centrally evoked contractions in the cat medial gastrocnemius muscle. II. Motoneuron firing-rate modulation. *J Neurophysiol*: 1996;75:38–50. https://doi.org/10.1152/jn.1996.75.1.38.

34. Prather JF, Clark BD, Cope TC. Firing rate modulation of motoneurons activated by cutaneous and muscle receptor afferents in the decerebrate cat. *J Neurophysiol*: 2002;88:1867–1879. https://doi.org/10.1152/jn.2002.88.4.1867.

35. Ranvier M. DEGENERATION AND REGENERATION OF DIVIDED NERVES. *The Journal of Nervous and Mental Disease*: 1874;1:93.

36. Kronecker H, Stirling W. The Genesis of Tetanus*. *The Journal of Physiology*: 1878;1:384–420. https://doi.org/10.1113/jphysiol.1878.sp000031.

37. Denny-Brown DE. The histological features of striped muscle in relation to its functional activity. *Proceedings of the Royal Society of London Series B, Containing Papers of a Biological Character*: 1929;104:371–411. https://doi.org/10.1098/rspb.1929.0014.

38. Burke RE. Motor unit types of cat triceps surae muscle. *J Physiol*: 1967;193:141–160. https://doi.org/10.1113/jphysiol.1967.sp008348.

39. Burke RE. Firing patterns of gastrocnemius motor units in the decerebrate cat. *J Physiol*: 1968;196:631–654. https://doi.org/10.1113/jphysiol.1968.sp008527.

40. Burke RE, Tsairis P. Anatomy and innervation ratios in motor units of cat gastrocnemius. *J Physiol*: 1973;234:749–765. https://doi.org/10.1113/jphysiol.1973.sp010370.

41. Burke RE, Levine DN, Tsairis P, Zajac FE. Physiological types and histochemical profiles in motor units of the cat gastrocnemius. *J Physiol*: 1973;234:723–748. https://doi.org/10.1113/jphysiol.1973.sp010369.

42. Burke RE, Levine DN, Salcman M, Tsairis P. Motor units in cat soleus muscle: physiological, histochemical and morphological characteristics. *J Physiol*: 1974;238:503–514. https://doi.org/10.1113/jphysiol.1974.sp010540.

43. Burke RE. Revisiting the notion of "motor unit types." *Prog Brain Res*: 1999;123:167–175.

44. Kernell D, Ducati A, Sjöholm H. Properties of motor units in the first deep lumbrical muscle of the cat's foot. *Brain Res*: 1975;98:37–55. https://doi.org/10.1016/0006-8993(75)90508-9.

45. Hammarberg C, Kellerth JO. Studies of some twitch and fatigue properties of different motor unit types in the ankle muscles of the adult cat. *Acta Physiol Scand*: 1975;95:231–242. https://doi.org/10.1111/j.1748-1716.1975.tb10047.x.

46. Dum RP, Kennedy TT. Physiological and histochemical characteristics of motor units in cat tibialis anterior and extensor digitorum longus muscles. *J Neurophysiol*: 1980;43:1615–1630. https://doi.org/10.1152/jn.1980.43.6.1615.

47. Zengel JE, Reid SA, Sypert GW, Munson JB. Membrane electrical properties and prediction of motor-unit type of medial gastrocnemius motoneurons in the cat. *J Neurophysiol*: 1985;53:1323–1344. https://doi.org/10.1152/jn.1985.53.5.1323.

48. Gardiner PF, Kernell D. The "fastness" of rat motoneurones: time-course of afterhyperpolarization in relation to axonal conduction velocity and muscle unit contractile speed. *Pflugers Arch*: 1990;415:762–766. https://doi.org/10.1007/BF02584018.

49. Gardiner PF. Physiological properties of motoneurons innervating different muscle unit types in rat gastrocnemius. *J Neurophysiol*: 1993;69:1160–1170. https://doi.org/10.1152/jn.1993.69.4.1160.

50. Botterman BR, Iwamoto GA, Gonyea WJ. Gradation of isometric tension by different activation rates in motor units of cat flexor carpi radialis muscle. *J Neurophysiol*: 1986;56:494–506. https://doi.org/10.1152/jn.1986.56.2.494.

51. Kernell D, Eerbeek O, Verhey BA. Relation between isometric force and stimulus rate in cat's hindlimb motor units of different twitch contraction time. *Exp Brain Res*: 1983;50:220–227. https://doi.org/10.1007/BF00239186.

52. Kernell D. Functional properties of spinal motoneurons and gradation of muscle force. *Adv Neurol*: 1983;39:213–226.

53. Farina D, Negro F, Muceli S, Enoka RM. Principles of Motor Unit Physiology Evolve With Advances in Technology. *Physiology (Bethesda)*: 2016;31:83–94. https://doi.org/10.1152/physiol.00040.2015.

54. Holobar A, Farina D, Gazzoni M, Merletti R, Zazula D. Estimating motor unit discharge patterns from high-density surface electromyogram. *Clin Neurophysiol*: 2009;120:551–562. https://doi.org/10.1016/j.clinph.2008.10.160.

55. Del Vecchio A, Holobar A, Falla D, Felici F, Enoka RM, Farina D. Tutorial: Analysis of motor unit discharge characteristics from high-density surface EMG signals. *Journal of Electromyography and Kinesiology*: 2020;53:102426. https://doi.org/10.1016/j.jelekin.2020.102426.

56. Negro F, Muceli S, Castronovo AM, Holobar A, Farina D. Multi-channel intramuscular and surface EMG decomposition by convolutive blind source separation. *J Neural Eng*: 2016;13:026027. https://doi.org/10.1088/1741-2560/13/2/026027.

57. Holobar A, Farina D. Blind source identification from the multichannel surface electromyogram. *Physiol Meas*: 2014;35:R143-165. https://doi.org/10.1088/0967-3334/35/7/R143.

58. Holobar A, Minetto MA, Botter A, Negro F, Farina D. Experimental Analysis of Accuracy in the Identification of Motor Unit Spike Trains From High-Density Surface EMG. *IEEE Transactions on Neural Systems and Rehabilitation Engineering*: 2010;18:221–229. https://doi.org/10.1109/TNSRE.2010.2041593.

59. Farina D, Holobar A, Merletti R, Enoka RM. Decoding the neural drive to muscles from the surface electromyogram. *Clinical Neurophysiology*: 2010;121:1616–1623. https://doi.org/10.1016/j.clinph.2009.10.040.

60. Merletti R, Holobar A, Farina D. Analysis of motor units with high-density surface electromyography. *Journal of Electromyography and Kinesiology*: 2008;18:879–890. https://doi.org/10.1016/j.jelekin.2008.09.002.

61. Holobar A, Zazula D. Correlation-based decomposition of surface electromyograms at low contraction forces. *Med Biol Eng Comput*: 2004;42:487–495. https://doi.org/10.1007/BF02350989.

62. A H, D Z. On the selection of the cost function for gradient-based decomposition of surface electromyograms. *Annual International Conference of the IEEE Engineering in Medicine and Biology Society IEEE Engineering in Medicine and Biology Society Annual International Conference*: 2008;2008. https://doi.org/10.1109/IEMBS.2008.4650254.

63. Enoka RM. Physiological validation of the decomposition of surface EMG signals. *J Electromyogr Kinesiol*: 2019;46:70–83. https://doi.org/10.1016/j.jelekin.2019.03.010.

64. Beauchamp JA, Khurram OU, Dewald JPA, Heckman CJ, Pearcey GEP. A computational approach for generating continuous estimates of motor unit discharge rates and visualizing population discharge characteristics. *J Neural Eng*: 2022;19:016007. https://doi.org/10.1088/1741-2552/ac4594.

65. Hassan AS, Fajardo ME, Cummings M, McPherson LM, Negro F, Dewald JPA, *et al.* Estimates of persistent inward currents are reduced in upper limb motor units of older adults. *J Physiol*: 2021;599:4865–4882. https://doi.org/10.1113/JP282063.

66. Kunugi S, Holobar A, Kodera T, Toyoda H, Watanabe K. Motor unit firing patterns on increasing force during force and position tasks. *J Neurophysiol*: 2021;126:1653–1659. https://doi.org/10.1152/jn.00299.2021.

67. Person RS, Kudina LP. Discharge frequency and discharge pattern of human motor units during voluntary contraction of muscle. *Electroencephalogr Clin Neurophysiol*: 1972;32:471–483. https://doi.org/10.1016/0013-4694(72)90058-2.

68. Monster AW, Chan H. Isometric force production by motor units of extensor digitorum communis muscle in man. *J Neurophysiol*: 1977;40:1432–1443. https://doi.org/10.1152/jn.1977.40.6.1432.

69. Powers RK, Heckman CJ. Synaptic control of the shape of the motoneuron pool input-output function. *J Neurophysiol*: 2017;117:1171–1184. https://doi.org/10.1152/jn.00850.2016.

70. Powers RK, Elbasiouny SM, Rymer WZ, Heckman CJ. Contribution of intrinsic properties and synaptic inputs to motoneuron discharge patterns: a simulation study. *J Neurophysiol*: 2012;107:808–823. https://doi.org/10.1152/jn.00510.2011.

71. Henneman E, Clamann HP, Gillies JD, Skinner RD. Rank order of motoneurons within a pool: law of combination. *J Neurophysiol*: 1974;37:1338–1349. https://doi.org/10.1152/jn.1974.37.6.1338.

72. Sharples SA, Miles GB. Maturation of persistent and hyperpolarization-activated inward currents shapes the differential activation of motoneuron subtypes during postnatal development. *Elife*: 2021;10:e71385. https://doi.org/10.7554/eLife.71385.

73. Kavanagh JJ, Taylor JL. Voluntary activation of muscle in humans: does serotonergic neuromodulation matter? *J Physiol*: July 2022. https://doi.org/10.1113/JP282565.

74. Lee RH, Heckman CJ. Enhancement of Bistability in Spinal Motoneurons In Vivo by the Noradrenergic α1 Agonist Methoxamine. *Journal of Neurophysiology*: 1999;81:2164–2174. https://doi.org/10.1152/jn.1999.81.5.2164.

75. Lee RH, Heckman CJ. Adjustable Amplification of Synaptic Input in the Dendrites of Spinal Motoneurons In Vivo. *J Neurosci*: 2000;20:6734–6740. https://doi.org/10.1523/JNEUROSCI.20-17-06734.2000.

76. Harvey PJ, Li X, Li Y, Bennett DJ. Endogenous Monoamine Receptor Activation Is Essential for Enabling Persistent Sodium Currents and Repetitive Firing in Rat Spinal Motoneurons. *Journal of Neurophysiology*: 2006;96:1171–1186. https://doi.org/10.1152/jn.00341.2006.

77. Hounsgaard J, Kiehn O. Ca++ dependent bistability induced by serotonin in spinal motoneurons. *Exp Brain Res*: 1985;57:422–425. https://doi.org/10.1007/BF00236551.

78. Hounsgaard J, Kiehn O. Serotonin-induced bistability of turtle motoneurones caused by a nifedipine-sensitive calcium plateau potential. *The Journal of Physiology*: 1989;414:265–282. https://doi.org/10.1113/jphysiol.1989.sp017687.

79. Lee RH, Heckman CJ. Influence of voltage-sensitive dendritic conductances on bistable firing and effective synaptic current in cat spinal motoneurons in vivo. *Journal of Neurophysiology*: 1996;76:2107–2110. https://doi.org/10.1152/jn.1996.76.3.2107.

80. Lee RH, Heckman CJ. Essential Role of a Fast Persistent Inward Current in Action Potential Initiation and Control of Rhythmic Firing. *Journal of Neurophysiology*: 2001;85:472–475. https://doi.org/10.1152/jn.2001.85.1.472.

81. Lee RH, Heckman CJ. Bistability in Spinal Motoneurons In Vivo: Systematic Variations in Persistent Inward Currents. *Journal of Neurophysiology*: 1998;80:583–593. https://doi.org/10.1152/jn.1998.80.2.583.

82. Jacobs BL, Martín-Cora FJ, Fornal CA. Activity of medullary serotonergic neurons in freely moving animals. *Brain Res Brain Res Rev*: 2002;40:45–52. https://doi.org/10.1016/s0165-0173(02)00187-x.

83. Veasey SC, Fornal CA, Metzler CW, Jacobs BL. Response of serotonergic caudal raphe neurons in relation to specific motor activities in freely moving cats. *J Neurosci*: 1995;15:5346–5359.

84. Enoka RM, Farina D. Force Steadiness: From Motor Units to Voluntary Actions. *Physiology (Bethesda)*: 2021;36:114–130. https://doi.org/10.1152/physiol.00027.2020.

85. Jones KE, Hamilton AF, Wolpert DM. Sources of signal-dependent noise during isometric force production. *J Neurophysiol*: 2002;88:1533–1544. https://doi.org/10.1152/jn.2002.88.3.1533.

86. Gobbo M, Cè E, Diemont B, Esposito F, Orizio C. Torque and surface mechanomyogram parallel reduction during fatiguing stimulation in human muscles. *Eur J Appl Physiol*: 2006;97:9–15. https://doi.org/10.1007/s00421-006-0134-8.